# The Death Spiral of T Pyxidis

Joseph Patterson,[1] Arto Oksanen,[2] Berto Monard,[3] Robert Rea,[4]
Franz-Josef Hambsch,[5] Jennie McCormick,[6] Peter Nelson,[7] Jonathan Kemp,[1]
William Allen,[8] Thomas Krajci,[9] Simon Lowther,[10] Shawn Dvorak,[11]
Thomas Richards,[12] Gordon Myers,[13] and Greg Bolt[14]

[1]*Department of Astronomy, Columbia University, 550 West 120th Street, New York, NY 10027, USA; jop@astro.columbia.edu, jonathan@astro.columbia.edu*

[2]*CBA–Finland, Verkkoniementie 30, FI–40950 Muurame, Finland; arto.oksanen@jklsirius.fi*

[3]*CBA–Kleinkaroo, PO Box 281, Calitzdorp 6660, South Africa; bmonard@mweb.co.za*

[4]*CBA–Nelson, Regent Lane Observatory, 8 Regent Lane, Richmond, Nelson 7020, New Zealand; reamarsh@ihug.co.nz*

[5]*CBA–Mol, Oude Bleken 12, B–2400 Mol, Belgium; hambsch@telenet.be*

[6]*CBA–Pakuranga, Farm Cove Observatory, 2/24 Rapallo Place, Farm Cove, Pakuranga, Auckland 2012, New Zealand; farmcoveobs@xtra.co.nz*

[7]*CBA–Victoria, Ellinbank Observatory, 1105 Hazeldean Road, Ellinbank 3821, Victoria, Australia; pnelson@dcsi.net.au*

[8]*CBA–Blenheim, Vintage Lane Observatory, 83 Vintage Lane, RD 3, Blenheim 7273, New Zealand; whallen@xtra.co.nz*

[9]*CBA–New Mexico, PO Box 1351 Cloudcroft, NM 88317, USA; tom_krajci@tularosa.net*

[10]*CBA–Pukekohe, Jim Lowther Observatory, 19 Cape Vista Crescent, Pukekohe 2120, New Zealand; simon@jlobservatory.com*

[11]*CBA–Orlando, Rolling Hills Observatory, 1643 Nightfall Drive, Clermont, FL 34711, USA; sdvorak@rollinghillsobs.org*

[12]*CBA–Melbourne, Pretty Hill Observatory, PO Box 323, Kangaroo Ground 3097, Victoria, Australia; tom@prettyhill.org*

[13]*CBA–San Mateo, 5 Inverness Way, Hillsborough, CA 94010, USA; gordonmyers@hotmail.com*

[14]*CBA–Perth, 295 Camberwarra Drive, Craigie, Western Australia 6025, Australia; gbolt@iinet.net.au*

**Abstract.**
    We report a long campaign to track the 1.8 hr photometric wave in the recurrent nova T Pyxidis, using the global telescope network of the Center for Backyard Astro-

physics. During 1996–2011, that wave was highly stable in amplitude and waveform, resembling the orbital wave commonly seen in supersoft binaries. The period, however, was found to increase on a timescale $\frac{P}{\dot{P}} = 3 \times 10^5$ yr. This suggests a mass transfer rate of $\sim 10^{-7}$ $M_\odot$/yr in quiescence. The orbital signal became vanishingly weak (< 0.003 mag) near maximum light of the 2011 eruption. After it returned to visibility near $V = 11$, the orbital period had increased by 0.0054(6) %. This is a measure of the mass ejected in the nova outburst. For a plausible choice of binary parameters, that mass is at least $3 \times 10^{-5}$ $M_\odot$, and probably more. This represents > 300 yr of accretion at the pre-outburst rate, but the time between outbursts was only 45 yr. Thus the erupting white dwarf seems to have ejected at least 6× more mass than it accreted. If this eruption is typical, the white dwarf must be eroding, rather than growing, in mass — dashing the star's hopes of ever becoming famous via a supernova explosion. Instead, it seems likely that the binary dynamics are basically a suicide pact between the eroding white dwarf and the low-mass secondary, excited and rapidly whittled down, probably by the white dwarf's EUV radiation.

## 1.  Introduction

T Pyxidis is the Galaxy's most famous recurrent nova. Six times since 1890, the star has erupted to $V = 6$, and then subsided back to quiescence near $V = 15$. With spectroscopy and detailed light curves known for most of these eruptions, and with a fairly bright quiescent counterpart, T Pyx has become a well-studied star — sometimes considered a prototype for recurrent novae. Selvelli et al. (2008) and Schaefer et al. (2010) give recent reviews.

Since they are believed (and in a few cases known) to possess massive white dwarfs accreting at a high rate, recurrent novae are a promising source for Type Ia supernovae. But since they also *eject* matter, their candidacy rests on the assumption that mass accretion in quiescence exceeds mass ejection in outburst. Estimates of these rates are notoriously uncertain, and that assumption has never undergone a significant test. A *dynamical* measure of the mass ejected, based on the precise orbital period change in outburst, would furnish by far the most precise and compelling evidence.

In the late 1980s, it was recognized that T Pyx might soon furnish that information, since an outburst was expected soon (1988, judging from the 1966 outburst and the 22-yr mean interval). However, the orbital period was not yet known; several photometric and spectroscopic studies gave discrepant periods, and all are now known to be incorrect. Schaefer et al. (1992, hereafter S92) identified a persistent photometric wave with a period of 0.076 d, but discounted that as a possible orbital period, since it did not appear to be coherent from month to month. They interpreted it as a "superhump" — arising from precession of the accretion disk — and estimated an underlying $P_{orb}$ near 0.073 d. A 1996–1997 observing campaign (Patterson et al. 1998, hereafter P98) revealed that the weak 0.076 d signal, difficult to discern over a single cycle, is actually quite coherent, maintaining a constant phase and waveform over many thousands of cycles. With a precise ephemeris, it bore all the earmarks of a *bona fide* orbital period. Remarkably, that study of all timings during 1986–1997 revealed an enormous rate of period increase, with $\frac{P}{\dot{P}} = 3 \times 10^5$ yr. Any remaining dissent from the orbital-period interpretation fell away when Uthas et al. (2010, hereafter UKS) found radial-velocity variations also following the 0.07622 d period, but only when the exact increasing-period photometric ephemeris was adopted (see their Figure 2).

This paper reports briefly on our long-term photometric study of T Pyx with the globally distributed telescopes of the Center for Backyard Astrophysics (CBA). All the "quiescent" data are basically consistent with the P98 ephemeris (slightly tweaked). And, as hoped, the signal returned after the 2011 eruption — with a different period. Thus the sought-after dynamical measure of ejected mass has become possible.

## 2. Observations in Quiescence (1996–2011)

Following the report of a persistent 1.8 hr quasiperiod by S92, we made T Pyx a priority target for time-series photometry. In the 1996–1997 campaign, we proved the existence of a strict 0.07622 d period, stable in phase and waveform over a 1-yr baseline — and deduced a long-term cycle count which tied together timings of minima over the full 1986–1997 baseline (P98). Some doubt still remained about this cycle count; it relied on quite sparse timings earlier than 1996, and also required hypothesizing a rate of (presumed orbital) period change which was orders of magnitude greater than anything previously seen in cataclysmic variables.

Great stability is the main credential certifying an orbital origin, and we have studied the light curves for stability and timing during each observing season since 1996. We accumulated ~ 1200 hr of time-series photometry. By 1999, it was clear that the main elements of the P98 study were confirmed. Averaged over each dense cluster of photometry during each season, the 1.8 hr signal was completely stable in period, waveform, amplitude, and phase. And the minima tracked the P98 ephemeris to high precision, thus verifying the cycle count and the signal's consequent very high $\dot{P}$.

Those timings of 1996–2011 minima, each averaged over typically 5–30 orbits, are rendered in the O–C diagram of Figure 1. The upward curve indicates a steadily increasing period, and the good fit of the parabola is consistent with a constant rate of period change. The curve corresponds to the ephemeris

$$\text{Minimum light} = \text{HJD } 2450124.831(1) + 0.0762263(2)\,E + 2.38(8) \times 10^{-11}\,E^2. \quad (1)$$

This implies $\frac{dP}{dt} = 6.4 \times 10^{-10}$, or $\frac{P}{\dot{P}} = 3.3 \times 10^5$ yr.

## 3. Observations After Eruption

The 2011 eruption was discovered and announced on 14 Apr, and we obtained time-series photometry on ~ 450 of the next ~ 600 nights, totalling ~ 2000 hr. We used the same techniques as at quiescence: segregate the data into dense clusters over ~ 10–30 nights, and look for periodic signals in each. Near maximum light, no periodic signals were found over the frequency range 3–1000 cycles/d. The (peak-to-trough) amplitude upper limit for signals near $\omega_{\text{orb}}$ was 3 – 5 mmag. The first obvious detection of a periodic signal occurred around day 170 ($V = 11$), when a 12-night time series yielded a clear signal at the orbital frequency, with an amplitude of 4 mmag. This signal grew steadily in amplitude as the star continued its decline from maximum light. Two dense clusters near day 70 ($V = 9$) also produced likely detections of $\omega_{\text{orb}}$. (These signals were not significant in the power spectra, but synchronous summations at $P_{\text{orb}}$ yielded the familiar waveform, and gave a timing of minimum light consistent with the post-eruption ephemeris.)

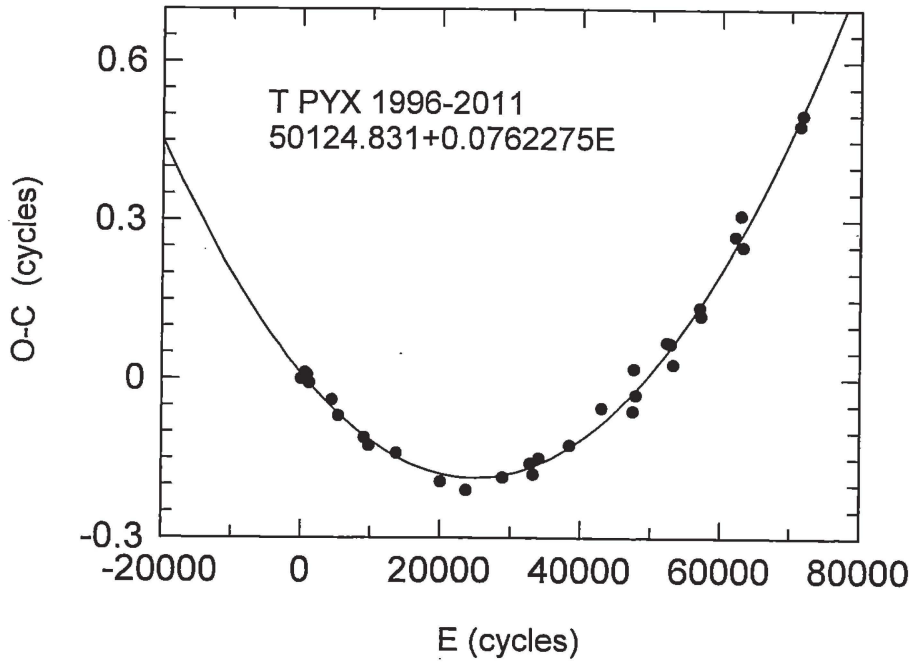

Figure 1.     O–C diagram of the timings of primary minima during 1996–2011, with respect to the test ephemeris shown in the figure. The fit to a parabola indicates acceptable representation with a constant rate of period change.

Each later segment showed a strong detection at $\omega_{orb}$, a weak detection at $2\omega_{orb}$, and no other signals. A typical power spectrum is shown in Figure 2, and is substantially identical to the power spectra of quiescence. For several segments the alias at $\omega_{orb}-3.00$ cycles/d is surprisingly strong, and we briefly considered whether that might be a detection of an independent signal (for which P98 suggested weak evidence, and which has been sometimes interpreted as evidence for magnetically channeled accretion). However, study of the spectral window showed that it is merely an alias. That particular alias is a special hazard of southern stars, since the southern planet is mostly water, with three major centers of astronomical research — Chile, South Africa, and Australia/New Zealand — spaced by $\sim 120°$ in longitude. (*Long* nightly time series are never fooled by this distant alias, but the necessarily-short runs near solar conjunction can be.)

For each segment we folded the $\sim 50$ orbits of data on $P_{orb}$, measured the averaged time of minimum light, and determined a best-fit period from the $\sim 30$ timings. That period is 0.0762336(1) d, an increase over the period just prior to eruption by 0.0054(5) %. So large a period change is very, very surprising: 7× larger than the $\Delta P$ predicted by Livio (1991) — and of the opposite sign!

Figure 3 shows the period changes since 1986. For 1996–2011, each point represents a 2-yr running average from the timings of minima. The 1986–1990 points are not certain, since they are mainly timings from single orbits (and thus subject to contamination by erratic flickering). Nevertheless, they agree with the 1996–2011 ephemeris extrapolated backwards in time, so we assume in Figure 3 that the full 1986–2011 cy-cle count is certain. Readers suspicious of this assumption should delete the first two points.

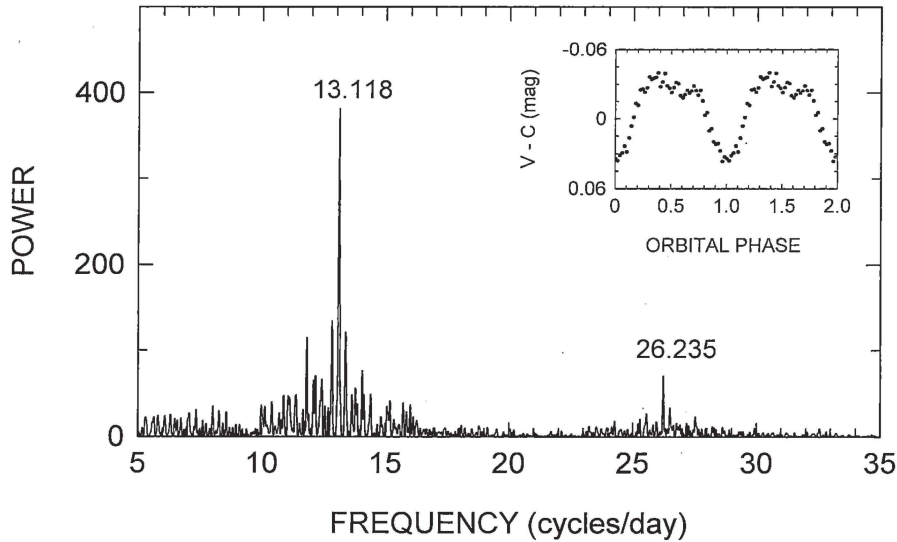

Figure 2. Power spectrum of a typical dense (20-d) segment of light curve after eruption. The only significant signals are $\omega_{orb}$ and its harmonics; upper limits at other frequencies are typically ~ 0.003 mag. Inset is the mean orbital light curve.

## 4. Interpretation

In quiescence, T Pyx's secondary transfers matter to the white dwarf — apparently through an accretion disk, since the emission lines show the doubled profiles characteristic of a disk (UKS). If total mass and angular momentum are conserved in this process, then the white dwarf gains matter at a rate

$$\dot{M}_1 = \frac{q\, M_1}{3\,(1-q)} \frac{\dot{P}}{P}, \qquad (2)$$

where $M_1$ is the white-dwarf mass and $q = \frac{M_2}{M_1}$. For our measured $\dot{P}$ and the binary parameters formally deduced by UKS ($M_1 = 0.7\, M_\odot$, $q = 0.2$), this implies $\dot{M}_1 = 1.8 \times 10^{-7}\, M_\odot$/yr. However, the line doubling and the photometric modulation are somewhat surprising if the binary inclination is as low as the UKS value ($10 \pm 2°$). We tend to favor a lower $q$ (closer to 0.1), which would raise $i$ to ~ 20° and bring $\dot{M}_1$ close to $1.0 \times 10^{-7}\, M_\odot$/yr.

This is also roughly the accretion rate implied from the luminosity. Correcting the P98 estimate for a distance of 4.8 Kpc (Nelson et al. 2012), we now estimate a bolometric luminosity of $1.1 \times 10^{36}$ erg/s. If this represents the energy of accretion onto the white dwarf, and if $m_1 = \frac{M_1}{1\, M_\odot}$, then the accretion rate is $6 \times 10^{-8}\, m_1^{1.8}\, M_\odot$/yr — or twice that if we count only the "disk" component. We average all four estimates to obtain $\dot{M}_1 = 1.2 \times 10^{-7}\, M_\odot$/yr.

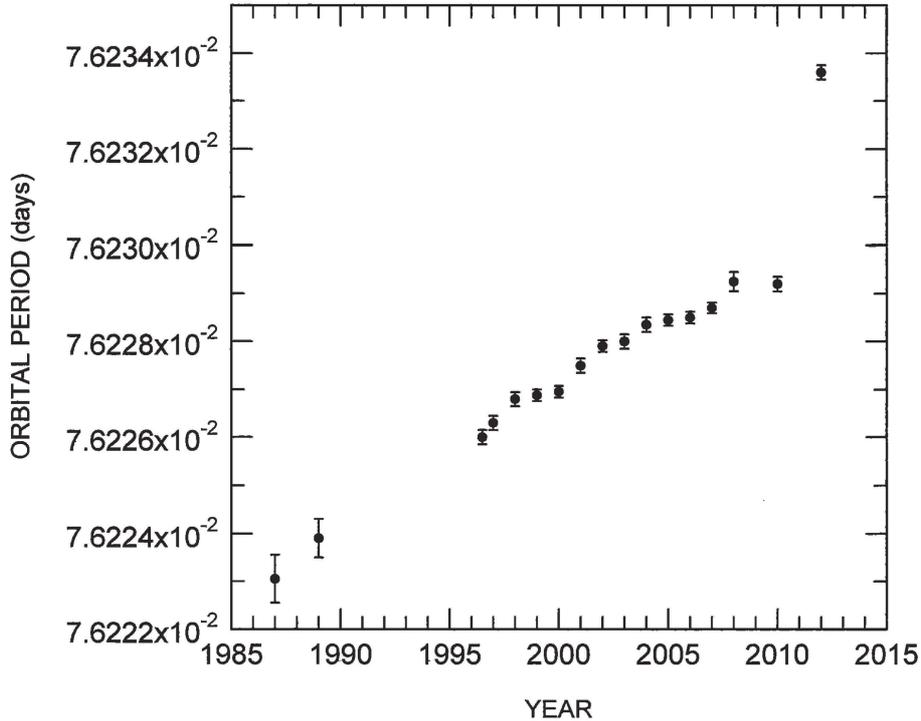

Figure 3.  The variation of $P_{orb}$ during 1986–2013. Each point represents a 2-yr running mean.

During eruption, mass loss should increase $P_{orb}$, and angular-momentum loss should decrease it. It's an open question which will dominate. But our observations show $\frac{\Delta P}{P} = +5.4 \times 10^{-5}$, indicating that mass loss wins. For the minimum plausible prescription for angular-momentum loss (radial ejection from the white dwarf), this implies a mass loss

$$\Delta M = 3.0 \times 10^{-5} \, m_1(1+q) \, M_\odot. \tag{3}$$

For $m_1 \approx 1$, this represents about 250 years of accretion, yet only 45 years elapsed since the 1966 outburst. So it appears that the white dwarf ejected at least ∼ 5–6 × more matter than it accreted.

One can nibble around the edges of this conclusion by revising some numbers ($m_1$, $q$, $i$, bolometric correction). But the assumption most susceptible to error is that the nova ejecta carry off very little angular momentum (just the specific angular momentum of the white dwarf). It's easy to imagine ways in which more angular momentum is carried away: from the secondary, from rotation, from frictional losses. But the observed $\Delta P$ is large, positive, and undeniable; so each of these would only *raise* $\Delta M$, strengthening the conclusion that the white dwarf erodes. We note that radio observations (from the free-free emission) also suggest a large $\Delta M$, probably near $10^{-4}$ $M_\odot$ (Nelson et al. 2012). Thus it now seems unlikely that the white dwarf in T Pyx — once considered a fine ancestor for a Type Ia supernova — will ever increase its mass at all, much less reach 1.4 $M_\odot$.

## 5. *Quo Vadis*, T Pyx?

We have now tracked the $P_{orb}$ evolution through 27 yr — just about the average interval between eruptions. The observations include an eruption, and the six known eruptions are pretty close counterparts, at least in their light curves. So with a little nip from Ockham's Razor, but without proof of course, it seems reasonable to consider the possibility that this evolution will continue: with $P_{orb}$ ever increasing, each nova event carrying off $\sim 10^{-4}$ $M_\odot$, and progressively whittling down the secondary to smaller mass (and probably larger radius, inflated by the continuing barrage of radiation on its surface). Knigge et al. (2000) present an elaboration of this possibility (though see Schaefer et al. 2010 for an alternative viewpoint).

We have always wondered why T Pyx is unique. This scenario offers a candidate explanation: because it is dying — annihilating its secondary in a paroxysm of repeated nova events, and lasting only $\sim 10^5$ more years (at the current rate). Some of the population statistics of cataclysmic variables (total space densities, ratio of long-period to short-period CVs) would make more sense if there were a way to kill off short-period CVs, thereby preventing them from swamping the local census (Patterson 1984, 1998). T Pyx may offer an embarrassingly gaudy but practical way to do this.

**Acknowledgments.** Thanks to the NSF for their grants AST12–11129 and AST09–08363 for support of this research, and to the AAVSO for their support and superb organization of variable-star astronomy over the past century. Some data were obtained at Cerro Tololo Interamerican Observatory, National Optical Astronomy Observatory, which is operated by AURA, under contract with the NSF. And a great deal of excellent data came from Caisey Harlingten's observatory in San Pedro de Atacama, Chile.